\documentclass[aps,prl,twocolumn,superscriptaddress]{revtex4-2}
\newcommand\fdg{\mbox{$.\!\!^\circ$}}
\usepackage{amsmath}
\usepackage{graphicx}
\usepackage{xcolor} 
\usepackage{aas_macros}

\begin{document}


\title{Relativistic effects of PSR~J1856--0039 double neutron star system in a 2.36-hour compact orbit}


\author{Z.~L. Yang}
\affiliation{National Astronomical Observatories, Chinese Academy of Sciences, Jia-20 Datun Road, ChaoYang District, Beijing 100012, China}
\affiliation{School of Astronomy and Space Science, University of Chinese Academy of Sciences, Beijing 100049, China}

\author{J.~L. Han}\email{Contact author: hjl@nao.cas.cn [ORCID: 0000-0002-9274-3092]}
\affiliation{National Astronomical Observatories, Chinese Academy of Sciences, Jia-20 Datun Road, ChaoYang District, Beijing 100012, China}
\affiliation{School of Astronomy and Space Science, University of Chinese Academy of Sciences, Beijing 100049, China}
\affiliation{State Key Laboratory of Radio Astronomy and Technology, Beijing 100101, China }

\author{W.~Q. Su}
\affiliation{National Astronomical Observatories, Chinese Academy of Sciences, Jia-20 Datun Road, ChaoYang District, Beijing 100012, China}
\affiliation{School of Astronomy and Space Science, University of Chinese Academy of Sciences, Beijing 100049, China}

\author{P.~F. Wang}
\affiliation{National Astronomical Observatories, Chinese Academy of Sciences, Jia-20 Datun Road, ChaoYang District, Beijing 100012, China}
\affiliation{School of Astronomy and Space Science, University of Chinese Academy of Sciences, Beijing 100049, China}
\affiliation{State Key Laboratory of Radio Astronomy and Technology, Beijing 100101, China }

\author{C. Wang}
\affiliation{National Astronomical Observatories, Chinese Academy of Sciences, Jia-20 Datun Road, ChaoYang District, Beijing 100012, China}
\affiliation{School of Astronomy and Space Science, University of Chinese Academy of Sciences, Beijing 100049, China}
\affiliation{State Key Laboratory of Radio Astronomy and Technology, Beijing 100101, China }

\author{T. Wang}
\affiliation{National Astronomical Observatories, Chinese Academy of Sciences, Jia-20 Datun Road, ChaoYang District, Beijing 100012, China}
\author{D.~J. Zhou}
\affiliation{National Astronomical Observatories, Chinese Academy of Sciences, Jia-20 Datun Road, ChaoYang District, Beijing 100012, China}

\author{Yi Yan}
\affiliation{National Astronomical Observatories, Chinese Academy of Sciences, Jia-20 Datun Road, ChaoYang District, Beijing 100012, China}
\affiliation{School of Astronomy and Space Science, University of Chinese Academy of Sciences, Beijing 100049, China}

\author{J. Xu}
\affiliation{National Astronomical Observatories, Chinese Academy of Sciences, Jia-20 Datun Road, ChaoYang District, Beijing 100012, China}
\affiliation{State Key Laboratory of Radio Astronomy and Technology, Beijing 100101, China }

\author{W.~C. Jing}
\affiliation{National Astronomical Observatories, Chinese Academy of Sciences, Jia-20 Datun Road, ChaoYang District, Beijing 100012, China}
\affiliation{School of Astronomy and Space Science, University of Chinese Academy of Sciences, Beijing 100049, China}

\author{N.~N. Cai}
\affiliation{National Astronomical Observatories, Chinese Academy of Sciences, Jia-20 Datun Road, ChaoYang District, Beijing 100012, China}

\author{R.~X. Xu}
\affiliation{Department of Astronomy, Peking University, Beijing 100871, China}

\author{H.~G. Wang} 
\affiliation{Department of Astronomy, School of Physics and Materials Science, Guangzhou University, Guangzhou 510006, Guangdong Province, China} 
\affiliation{National Astronomical Data Center, Great Bay Area, Guangzhou 510006, Guangdong Province, China}

\author{X.~P. You}
\affiliation{School of Physical Science and Technology, Southwest University, Chongqing 400715, China}


\date{\today}

\begin{abstract}
Compact double neutron star (DNS) systems are unique laboratories for testing gravitational theories and studying DNS mergers. Here we report the properties of a new DNS system, PSR J1856--0039, discovered in the Five-hundred-meter Aperture Spherical radio Telescope (FAST). The pulsar is mildly recycled with a period of 23.4~ms in a compact eccentric orbit ($e=0.106$) with an orbital period of 2.36 hours. By following up FAST observations, we measured the relativistic effects, including the orbital period derivative $\dot{P}_{\rm orb}=-1.284\pm0.019\times10^{-12}$~s\,s$^{-1}$, periastron advance $\dot\omega=17.5859\pm0.0007$~deg\,yr$^{-1}$, and Einstein delay $\gamma=0.445\pm0.011$~ms. This DNS system has a low orbital inclination of $i=133\fdg2\pm1\fdg1$ and the lowest total mass of any known DNS, $M_{\rm tot}=2.48841\pm0.00015~M_\odot$, with a determined pulsar mass of $1.304\pm0.022~M_\odot$ and a companion mass of $1.185\pm0.022~M_\odot$, one of the lowest neutron-star masses. The observed orbital decay due to gravitational-wave emission $\dot{P}^{\rm GW}_{\rm orb,obs}$ and the orbital decay predicted by general relativity $\dot{P}^{\rm GW}_{\rm orb,pred}$ 
are consistent at a level of $\dot{P}^{\rm GW}_{\rm orb,obs}/\dot{P}^{\rm GW}_{\rm orb,pred}=$1.009(14) (68\% confidence). This DNS will merge after 82 Myr and may form a stable neutron star or collapse into a black hole after spin-down. Long-term monitoring could potentially probe the Lense‑Thirring precession.
\end{abstract}

\maketitle

\section{Introduction} 

Among all known pulsars, approximately 30 are hosted in double neutron star (DNS) systems \cite{Manchester+2005AJ....129.1993M}. Two neutron stars in compact binaries  --- each possessing an extremely strong gravitational field and high binding energy --- orbit around their common center of mass with orbital periods ranging from a few hours to several tens of days. They serve as unique laboratories for testing predictions of gravitational theories. For example, the first known binary pulsar PSR B1913+16 is in a DNS system with an orbital period of 7.75 hours \cite{Hulse+1975ApJ...195L..51H}. Long-term timing of this pulsar has revealed a decay in its orbital period, consistent with orbital damping due to gravitational-wave emission as predicted by general relativity \cite{Taylor+1979Natur.277..437T, Weisberg+2016ApJ...829...55W}, providing the first indirect evidence for the existence of gravitational waves. Such orbital decay has also been measured with high precision in the double pulsar system, PSR J0737--3039A/B, which has an orbital period of 2.45 hours. The derived orbital decay $\dot{P}^{\rm GW}_{\rm orb,obs}$ caused by gravitational-wave emission is in perfect agreement with the prediction of general relativity $\dot{P}^{\rm GW}_{\rm orb,pred}$, with $\dot{P}^{\rm GW}_{\rm orb,obs}/\dot{P}^{\rm GW}_{\rm orb,pred}=0.99996\pm0.00006$, confirming general relativity to a relative accuracy of $1.3\times10^{-4}$ at 95\% confidence \cite{Kramer+2021PhRvX..11d1050K}. The most compact DNS system, PSR J1946+2052, has provided the second-most precise test of the gravitational-wave emission, with 
$\dot{P}^{\rm GW}_{\rm orb,obs}/\dot{P}^{\rm GW}_{\rm orb,pred}= 
1.00005\pm0.00091$, achieved just 7 years after its discovery \cite{Meng+2025AA...704A.153M}. 
The continuous emission of gravitational waves will ultimately lead to the merger of a DNS system if it has a sufficiently short orbital period or a sufficiently high eccentricity. Such mergers produce powerful gravitational-wave signals, as GW170817 \cite{Abbott+2017PhRvL.119p1101A} and GW190425 \cite{Abbott+2020ApJ...892L...3A} were detected by LIGO and Virgo.
These DNS systems also provide not only landmark results for gravity tests, but also valuable constraints on the Galactic DNS merger rate \cite{Burgay+2003Natur.426..531B, Pol+2019ApJ...870...71P}.
The determined total masses and mass ratios of DNS mergers are directly linked to the nature of the merger remnant and the efficiency of heavy $r$-process nucleosynthesis \cite{Sarin+2021GReGr..53...59S}.

\begin{table}[bt]
\centering
\footnotesize
\caption{FAST observations of PSR~J1856--0039. Listed are the observation date, FAST project number, project principal investigator (PI),  FAST beam name, and observation time in minutes.}
\footnotesize
\setlength{\tabcolsep}{5pt}
    \begin{tabular}{cllr}
    \hline
         Obs. Date & FAST Beam name & Obs. mode & Obs. time \\ 
     (yyyymmdd)     &      &  & (minutes)        \\
    \hline
    20200504 & G33.18--1.44P4M14  & snapshot & 5   \\
    20211227 & G32.79--1.61P1M19  & snapshot & 5   \\
    20220116 & J185605--004136M08 & tracking & 15  \\
    20220130 & J185558--003916M08 & tracking & 15  \\
    20220711 & G33.18--1.44P4M14  & snapshot & 5   \\
    20221201 & J1856--0040M01     & tracking & 150 \\
    20230321 & J185651--003629M04 & tracking & 15  \\
    20230808 & J1856--0040M01     & tracking & 72  \\
    20231117 & J1856--0040M01     & tracking & 148 \\
    20240118 & J1856--0041M08     & tracking & 5   \\
    20240215 & J1856--0040M01     & tracking & 72  \\
    20240404 & J1856--0040M01     & tracking & 135 \\
    20240817 & J1856--0040gM01    & tracking & 93  \\
    20241019 & J1856--0040gM01    & tracking & 120 \\
    20241209 & J1856--0040gM01    & tracking & 120 \\
    20250305 & J1856--0040gM01    & tracking & 120 \\
    20250607 & J1856--0040gM01    & tracking & 144 \\
    \hline
    \end{tabular}
    \label{obsinfo}
\end{table}

Owing to these important scientific applications \cite{Freire+2024LRR....27....5F}, the discovery of compact DNS systems has been a key motivation for numerous pulsar surveys. %
Here we report the discovery of a compact DNS system, PSR J1856--0039, from the Five-hundred-meter Aperture Spherical radio Telescope (FAST) \cite{Nan+2006ScChG..49..129N, Nan+2011IJMPD..20..989N} Galactic Plane Pulsar Snapshot (GPPS) survey \cite{Han+2021RAA....21..107H, Han+2025RAA}. This survey aims to discover pulsars in the Galactic plane within FAST's visible sky and has so far detected over 850 new pulsars, including four published DNS systems: PSRs J0528+3529, J0641+0448, J1844--0128, and J1901+0658 \cite{Su+2024MNRAS, Wang+2025RAA, Yang+2026ApJ..1000L...1Y}. PSR J1856--0039 is the fifth DNS system discovered in the survey. It has a short orbital period of 2.36 hours and a mild eccentricity of 0.106. We have obtained a phase-connected timing solution that yields precise measurements of the neutron star masses, one of which is the second-lowest on record. The system is well described by general relativity, with prospects for future measurements of Lense-Thirring precession.

\begin{figure}[bt]
    \centering
    \includegraphics[width=0.95\columnwidth]{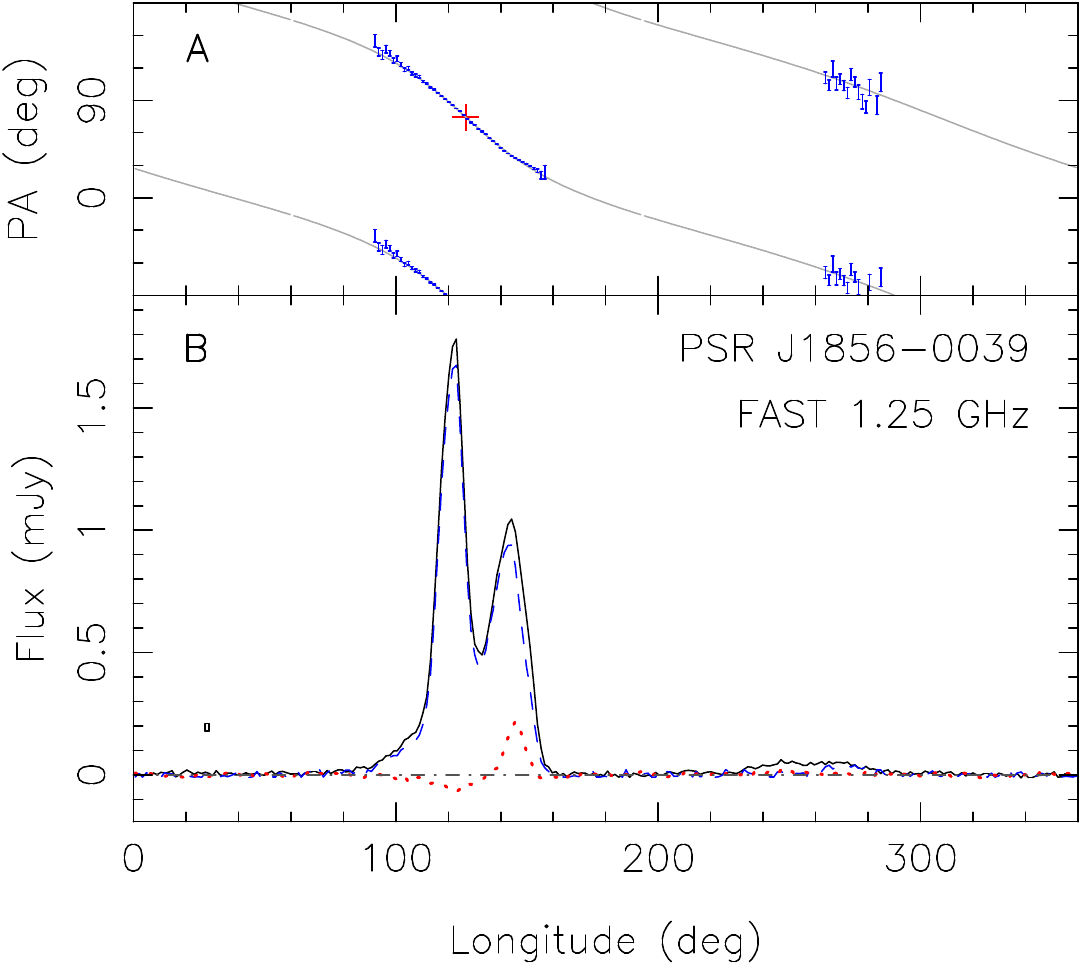}
    \caption{Polarization profiles of PSR J1856--0039 observed by FAST at 1.25 GHz integrated over 13 hours. Subpanel (A) shows the position angles (PAs) of linearly polarized emission at infinite frequency. The error bars indicate $\pm1\sigma$ uncertainties. The gray dotted line shows the best fit to the PA data with the RVM, with $\alpha = 114\fdg5 \pm 2\fdg6$ and $\zeta = 139\fdg1 \pm 3\fdg2$. The red cross marks the point of maximum PA gradient at the longitude of $126^\circ.7\pm0^\circ.3$ with the PA of $74^\circ.3\pm0^\circ.7$. Subpanel (B) shows the total intensity profile $I$ (solid line), linear $L$ (dashed line) and circular $V$ (dotted line, positive for the left-hand sense) polarization profiles, with a small square rectangle in the bottom left corner denoting $\pm2\sigma_{\rm I,off}$ height and 1-bin width.}
    \label{pol}
\end{figure}

\section{\label{sec2}FAST Observations and Data Reduction}

To date, PSR~J1856--0039 has been detected in a total of 17 FAST observational sessions, as listed in Table~\ref{obsinfo}. All FAST observations used the $L$-band 19-beam receiver, covering a frequency range from 1.0 to 1.5 GHz with a sampling time of 49.152~{\textmu}s. The three snapshot observations were recorded with 2048 frequency channels and 2 polarization channels ($XX$ and $YY$), while the tracking observations utilized 2048 frequency channels and 4 polarization channels ($XX$, $YY$, $\text{Re}[XY]$, and $\text{Im}[XY]$). As part of the standard calibration procedure, the modulated calibration signals with a period of 2.01326~s and an amplitude of 1.1~K were injected into the feed at either the beginning or the end of each FAST observation session \cite{Han+2021RAA....21..107H}.

PSR~J1856--0039 was first detected in a snapshot observation by FAST on 2020 May 4. After that, PSR~J1856--0039 was serendipitously detected in several subsequent observations (see Table~\ref{obsinfo}). Using these observations, we found its orbital parameters following the method proposed by \citet{Freire+2001MNRAS.322..885F} and   \citet{Bhattacharyya+2008MNRAS.387..273B}. The fitting results indicated that PSR~J1856--0039 is most likely in a compact DNS system with orbital eccentricity $e=0.106$ and orbital period $P_{\rm orb}=2.36$ hr.  
We conducted a 2.5-hour tracking observation on 2022 December 1 to fully cover its orbit (see Table~\ref{obsinfo}), and confirmed the orbital period.

\begin{table}
	\centering
	\caption{Timing model parameters and derived parameters for PSR J1856$-$0039.}
    \footnotesize
	\begin{tabular}{lr} 
		\hline
        \multicolumn{2}{c}{General information} \\
        \hline         
		PSRJ                &    J1856 -- 0039 \\	
		Number of TOAs      &    253 \\
		MJD range           &    58972 to 60832 \\
        MJD of period determination&  59900 \\
        UNITS       &   TDB  \\
        Solar System ephemeris   &    DE440 \\
        EFAC   &   0.79 \\
        \hline  
         \multicolumn{2}{c}{Model Parameter}  \\
	 	\hline
        R.A. (J2000)  &   $18^{\rm h}56^{\rm m}48^{\rm s}.91130(6)^\dagger$\\
		decl. (J2000)  &  -$00^\circ39'38''.0600(16)$  \\
		Spin frequency, $\nu$ (Hz) &   42.750461065640(3)  \\ 
		Spin frequency derivative, $\dot{\nu}$ (Hz $\rm s^{-1}$) & -1.19379(6)$\times10^{-15}$\\ 
		Dispersion measure, DM (pc\, cm$^{-3}$)    & 43.1250(8)\\
        Faraday rotation measure, RM (rad\,m$^{-2}$) & -24.5(3) \\
        Proper motion in R.A., $\mu_\alpha$ (mas yr$^{-1}$) & -$1.5(5)$ \\
        Proper motion in decl., $\mu_\delta$ (mas yr$^{-1}$)& -$6.1(9)$ \\
		Orbital period, $P_{\rm orb}$ (days)  & 0.09851259274(5) \\
		Projected semi-major axis, $x$ (lt-s) & 0.979262(11) \\
		Orbital eccentricity, $e$  &  0.106322(11)  \\
        Longitude of periastron, $\omega$ (deg) & -1.7167(6) \\
        Epoch of periastron, $T0$ (MJD) &  60444.14781000(16) \\
        Orbital period derivative, $\dot{P}_{\rm orb}$ (s\,s$^{-1}$) & -1.284(19)$\times10^{-12}$ \\
        Rate of periastron advance, $\dot{\omega}$ (deg\,yr$^{-1}$) & 17.5859(7) \\
        Einstein delay, $\gamma$ (ms) & 0.445(11)  \\
         
		\hline
        \multicolumn{2}{c}{Derived Parameter}  \\
	 	\hline
		Galactic longitude, $l$ (deg)  & 32.956880  \\
		Galactic latitude, $b$ (deg)   & -1.497483   \\
        DM Distance, $d_{\rm NE2025}$ (kpc)       &   2.0 \\
        DM Distance, $d_{\rm YMW16}$ (kpc)       &   1.3 \\
        Spin period, $P$ (s)  & 0.0233915605837462(15) \\
		Spin period derivative, $\dot{P}$ (s\,s$^{-1}$) &6.5320(3)$\times10^{-19}$ \\
		Characteristic age, $\tau_{\rm c}$ (Myr)  & 568  \\
		Surface magnetic field, $B_{\rm s}$ ($10^{9}$G)  & 3.96 \\
		Mass function (M$_\odot$)   &  0.103895(4) \\
        Total mass ($\rm M_\odot$)  &  2.48841(15) \\
        Companion mass$^{\dagger\dagger}$, $M_{\rm c}$ (M$_\odot$) & 1.185(22) \\
        Pulsar mass$^{\dagger\dagger}$, $M_{\rm p}$ (M$_\odot$) & 1.304(22) \\
        Excess $\dot{P}_{\rm orb}$$^{\dagger\dagger}$, $\dot{P}^{\rm GW}_{\rm orb,obs}-\dot{P}^{\rm GW}_{\rm orb,pred}$ (s\,s$^{-1}$) &  $-1.2(18)\times10^{-14}$ \\
        Predicted $\dot{P}_{\rm orb}^{\rm GW}$, $\dot{P}^{\rm GW}_{\rm orb,pred}$ (s\,s$^{-1}$) & -1.27248(13)$\times10^{-12}$ \\ 
        Coalescence timescale, $\tau_{\rm GW}$ (Myr) & 82  \\
		\hline
    \end{tabular}
    \label{timing_model}\\
    $^\dagger$ Numbers in parentheses represent 1$\sigma$ errors in the final decimal.\\
    $^{\dagger\dagger}$ $M_{\rm c}$, $M_{\rm p}$, and excess $\dot{P}_{\rm orb}$ are fitted using DDGR model.
\end{table}

\begin{figure}
    \centering
    \includegraphics[width=0.9\columnwidth]{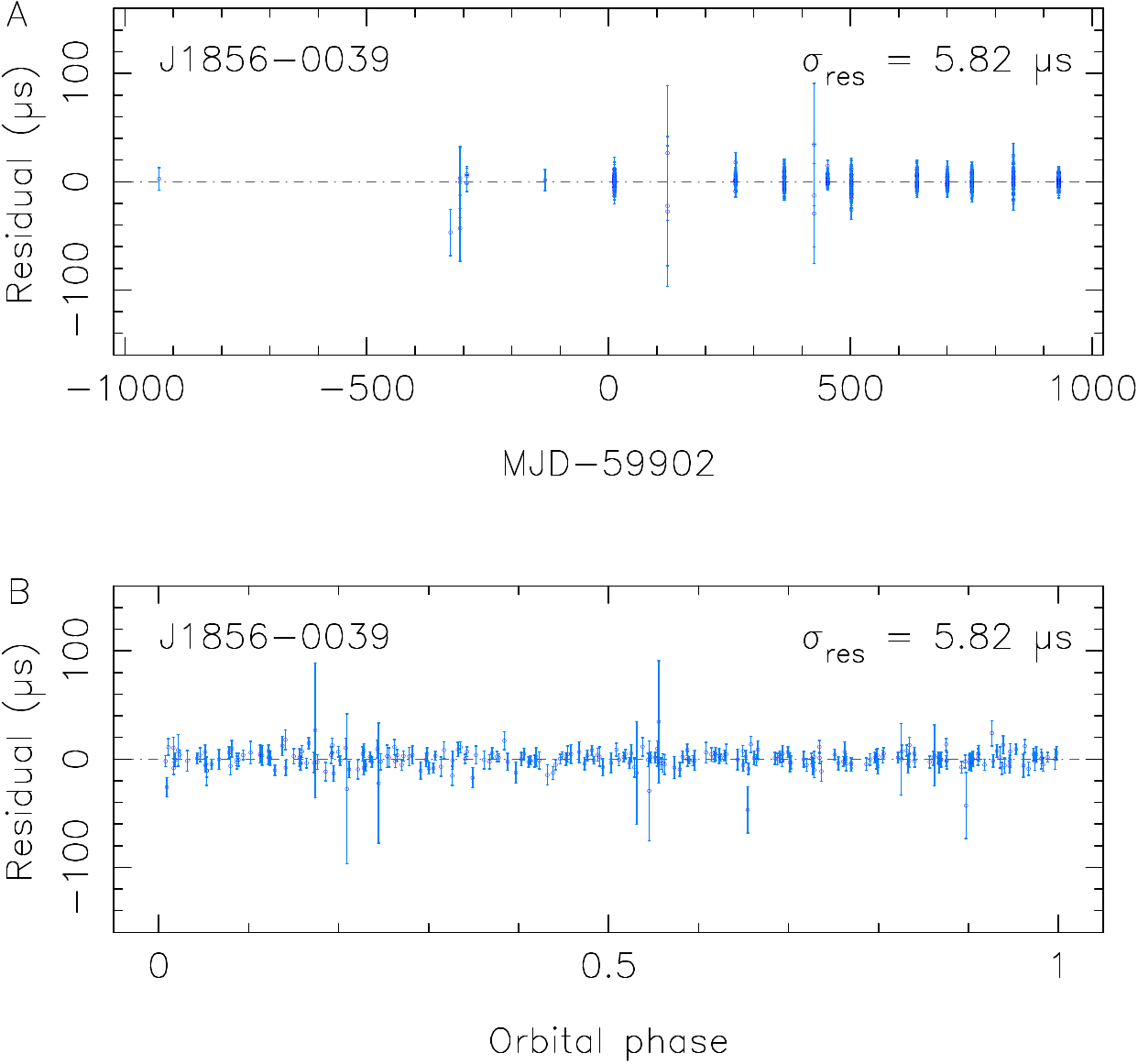}
    \caption{Timing residuals of PSR~J1856--0039. The weighted RMS of timing residuals is 5.82~{\textmu}s. Panels A and B show the timing residuals as a function of MJD and orbital phase, respectively. 
    }
    \label{timing_residuals}
\end{figure}


Using initial orbital parameters, we folded the raw \textsc{psrfits} data into pulse profiles with \textsc{dspsr} \cite{Straten+2011PASA...28....1V}. Profiles were calibrated and cleaned with \textsc{psrchive} \cite{Hotan+2004PASA...21..302H}. The FAST gain of 16.1~K\,Jy$^{-1}$ and $T_{\rm cal}$ of 1.1~K  \cite{Jiang+2020RAA....20...64J} have been used to estimate flux densities. 

To reduce data volume, we averaged frequency channels into four sub-bands and sub-integrations into 300 s using \texttt{pam}. Compared to a 1D template from \texttt{paas}, TOAs were extracted with \texttt{pat}. The DM was refined by comparing TOAs from the four sub-bands. Observational data were then integrated over all frequency channels using the refined DM value, from which we extracted new TOAs for timing analysis using \textsc{tempo2} \cite{Hobbs+2006MNRAS.369..655H}. A phase-connected solution for PSR~J1856--0039 was obtained following the procedures in \cite{Freire+2018MNRAS.476.4794F}, and then it was used to refold the data.

For polarization profiles, we analyzed the 144-min tracking observation on 2025 June 07. Using \texttt{rmfit}, we measured a Faraday rotation measure $RM_{\rm obs}=-22.52\pm0.11$ rad m$^{-2}$. The ionospheric contribution ($2.0\pm0.3$ rad m$^{-2}$) was determined using the codes for vertical \textsc{TEC maps}, \textsc{IGRF-13}, and \textsc{IONFR} \cite{Sotomayor+2013A&A...552A..58S}, yielding an intrinsic $RM=-24.5\pm0.3$ rad m$^{-2}$. After correcting for Faraday rotation in all tracking observations using the center beam (M01) of the 19-beam L-band receiver, we found no significant changes in either the polarization profiles \ or the position angles. Therefore, all FAST polarization observations were then combined using \texttt{psradd} to get the averaged polarization profiles, as shown in Fig.~\ref{pol}. 

Using \texttt{psrsmooth}, we created a new template from the averaged polarization profiles. To improve timing precision, we applied the Matrix Template Matching technique \cite{van+2006ApJ...642.1004V} to extract TOAs from the tracking observations.
Because the signals from PSR~J1856--0039 are highly polarized, following the procedures in \cite{van+2006ApJ...642.1004V}, the calibrated FAST observation data with 4 polarization channels can achieve greater precision of TOAs than uncalibrated data. We then iterated the timing analysis by using the DD binary model \cite{Damour+1985AIHPA..43..107D, Damour+1986AIHS...44..263D}, in which relativistic effects are described by post-Keplerian parameters. The Solar System ephemeris used is DE440 \cite{Park+2021AJ....161..105P}.
The final timing model and residuals are presented in Table~\ref{timing_model} and Fig.~\ref{timing_residuals}, respectively.

\section{\label{sec3} Relativistic effects and mass measurements}

PSR J1856--0039 is a mildly recycled pulsar with a spin period of 23.4 ms and a surface magnetic field of $4.0\times10^9$~G, suggesting that its companion had experienced the extreme stripping phase during the Case BB Roche-lobe overflow of the binary evolution \cite{Tauris+2017ApJ...846..170T}. 

The mass function $f(M_{\rm p},M_{\rm c},i)$ derived from observations is given by
\begin{equation}
    f(M_{\rm p},M_{\rm c},i)=\frac{(M_{\rm c} \sin i)^3}{(M_{\rm p}+M_{\rm c})^2}=\frac{4\pi^2}{G}\frac{x^3}{P_{\rm orb}^2}=0.1039~M_\odot,
    \label{fn}
\end{equation}
where $M_{\rm p}$ is the pulsar mass, $M_{\rm c}$ is the companion mass, $i$ is the orbital inclination (the angle between the observers' line of sight and the binary's orbital angular momentum), $G$ is the gravitational constant, and $x$ is the projected semi-major axis. Based on our FAST observations, we have also measured three post-Keplerian parameters: $\dot{\omega}=17.5859\pm0.0007$ deg\,yr$^{-1}$, $\dot{P}_{\rm orb}=(-1.284\pm0.019)\times10^{-12}$~s\,s$^{-1}$, and $\gamma=0.445\pm0.011$~ms which are used below for gravity tests and mass measurements. 

\begin{figure}
\centering
\includegraphics[width=0.95\columnwidth]{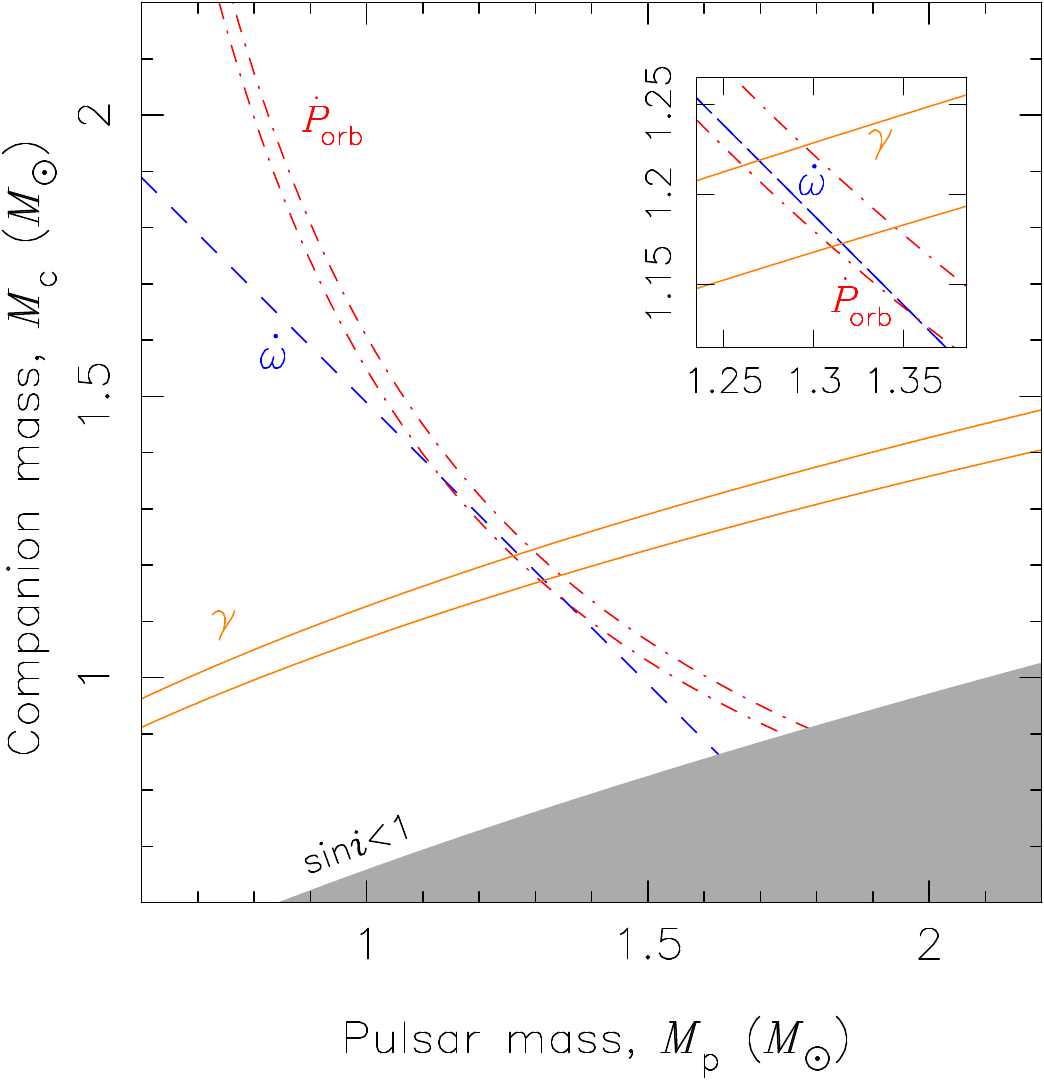} 
\caption{The mass–mass diagram for the PSR J1856--0039 DNS system. The gray region is not allowed by the mass function (for $\sin i \leq 1$). The constraints on possible NS masses by the $1\sigma$ uncertainties of three post-Keplerian parameters $\dot{\omega}$, $\dot{P}_{\rm orb}$, and $\gamma$, which are plotted under the assumption that general relativity is correct. The uncertainty in the $\dot{\omega}$ region is thinner than the line in the figure. The constrained regions intersect in a small section (see the enlarged view in the upper right corner),  confirming that GR is sufficient. 
}
   \label{m1m2_gpps0347}
\end{figure}

\begin{figure}
\centering
\includegraphics[width=0.95\columnwidth]{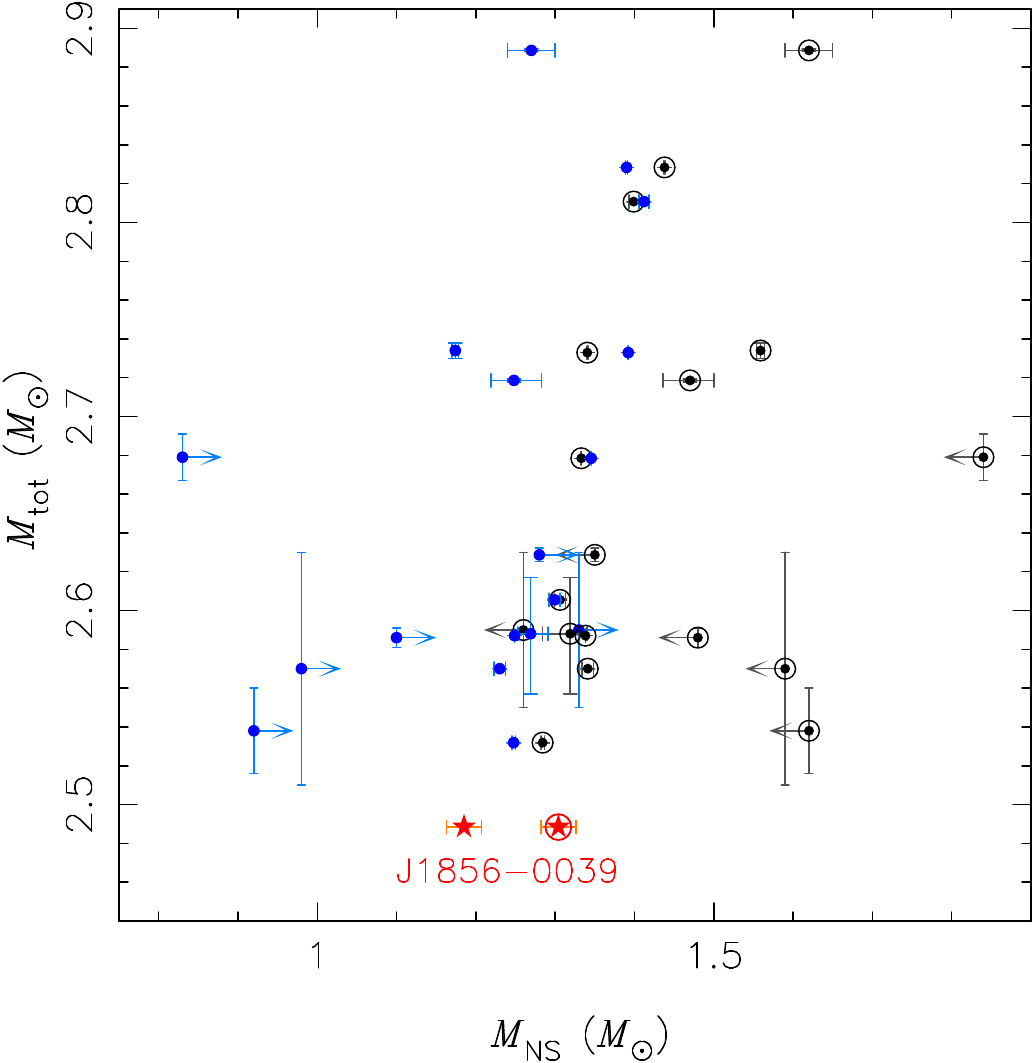}
\caption{The total masses of DNS systems versus individual neutron star masses. The masses of confirmed DNS systems with precise total mass measurements ($M_{\rm tot}$ uncertainty less than 0.1~$M_\odot$) are taken from literature \cite{Ferdman+2014MNRAS.443.2183F, Fonseca+2014ApJ...787...82F, Martinez+2015ApJ...812..143M, Swiggum+2015ApJ...805..156S, Weisberg+2016ApJ...829...55W, Martinez+2017ApJ...851L..29M, Ferdman+2020Natur.583..211F, Haniewicz+2021MNRAS.500.4620H, Kramer+2021PhRvX..11d1050K, Sengar+2022MNRAS.512.5782S, Cameron+2023MNRAS.523.5064C, Colom+2023A&A...678A.187C, McEwen+2024ApJ...962..167M, Su+2024MNRAS, Tan+2024ApJ...966...26T, Zhao+2024ApJ...964L...7Z, Meng+2025AA...704A.153M, Yang+2026ApJ..1000L...1Y}  and plotted. The first-born neutron stars in these systems are generally recycled and marked by black circles, while second-born neutron stars are denoted by dots. PSR J1856--0039 and its companion neutron star are indicated by the two stars, with the circle for the millisecond pulsar. 
}
\label{Mtot-Mns}
\end{figure}


To use post-Keplerian parameters for studying relativistic effects, kinematic corrections for $\dot{P}_{\rm orb}$ and $\dot{\omega}$ must be considered if their contributions are comparable to or larger than the measurement uncertainties.
The observed orbital period derivative ${\dot{P}}_{\rm{orb}}$ can be decomposed as:
\begin{equation}
{\dot{P}}_{\rm{orb}}={\dot{P}}_{\rm{orb,obs}}^{\rm{GW}}+{\dot{P}}_{\rm{orb}}^{\rm{Shk}}+{\dot{P}}_{\rm{orb}}^{\rm{Gal}},
\end{equation}
where ${\dot{P}}_{\rm{orb}}^{\rm{Shk}}$ represents the contribution from proper motion \cite[Shklovskii effect;][]{Shklovskii+1970SvA....13..562S}, and ${\dot{P}}_{\rm{orb}}^{\rm{Gal}}$ arises from the Galactic gravitational potential. The distance of PSR J1856--0039 estimated from the Galactic electron-density models NE2025 \cite{NE2025} and YMW16 \cite{ymw16} is 2.0~kpc or 1.3~kpc, respectively. Based on these values, the resulting total correction to ${\dot{P}}_{\rm{orb}}$ is approximately $1 \times 10^{-15}$ s\,s$^{-1}$, which is negligible for the current analysis. The contribution of proper motion to the periastron advance rate is of order $10^{-6}$ deg\,yr$^{-1}$, much smaller than its uncertainty $\sigma_{\dot{\omega}}$. We therefore conclude that no corrections to the measured post-Keplerian parameters are required at the present precision.


In general relativity, the post-Keplerian parameters $\dot{\omega}$, $\dot{P}_{\rm orb}^{\rm GW}$, and $\gamma$ can be expressed as functions of the pulsar and companion masses (the leading-order expression):
\begin{equation}
\begin{cases}
\dot{\omega}=3\left(\dfrac{P_{\rm orb}}{2\pi}\right)^{-5/3}\left[G(M_{\rm p}+M_{\rm c})\right]^{2/3}c^{-2}(1-e^2)^{-1},\\[1.2ex]
\dot{P}^{\rm GW}_{\rm orb,pred}=-\dfrac{192\pi}{5}\left(\dfrac{2\pi G}{c^3 P_{\rm orb}}\right)^{5/3}\dfrac{M_{\rm p}M_{\rm c}}{(M_{\rm p}+M_{\rm c})^{1/3}}f(e),\\[1.2ex]
\gamma=e\left(\dfrac{P_{\rm orb}}{2\pi}\right)^{1/3}G^{2/3}\dfrac{M_{\rm c}(M_{\rm p}+2M_{\rm c})}{(M_{\rm p}+M_{\rm c})^{4/3}}c^{-2},
\end{cases}
\end{equation}
where $f(e)=(1+\frac{73}{24}e^2+\frac{37}{96}e^4)(1-e^2)^{-7/2}$ and $c$ is the speed of light. Fig.~\ref{m1m2_gpps0347} shows the constraints on the NS masses of PSR J1856--0039 derived from the measured post-Keplerian parameters under the assumption of general relativity. Within $\pm1\sigma$ uncertainties, the corresponding NS masses of the three post-Keplerian parameters are self-consistent, indicating that general relativity passes the test.  $\dot{P}^{\rm GW}_{\rm orb,obs}-\dot{P}^{\rm GW}_{\rm orb,pred}$ is constrained to $(-1.2\pm1.8)\times10^{-14}$~s\,s$^{-1}$ using the DGR binary model, which assumes the validity of general relativity \cite{Taylor+1987grg..conf..209T, Taylor+1989ApJ...345..434T}. This result validates general relativity at a level of $\dot{P}^{\rm GW}_{\rm orb,obs}/\dot{P}^{\rm GW}_{\rm orb,pred}=1.009\pm0.014$. Future longer-term observations of the PSR J1856--0039 binary can provide more stringent tests of gravitational theories.

We applied the DDGR binary model \cite{Taylor+1987grg..conf..209T, Taylor+1989ApJ...345..434T} to constrain the masses of PSR J1856--0039 and its neutron star companion\footnote{The DDGR model only accounts for the leading-order terms of the post-Keplerian parameters, which is currently enough accurate because $\dot\omega_{\rm LT}$ and $\dot{\omega}_{\rm 2PN}$ cancel out each other ($\dot\omega_{\rm LT}+\dot{\omega}_{\rm 2PN}\ll\sigma_{\dot{\omega}}$).}, yielding $M_{\rm tot}=2.48841\pm0.00015~M_\odot$, $M_{\rm p}=1.304\pm0.022~M_\odot$, and $M_{\rm c}=1.185\pm0.022~M_\odot$. In contrast to other confirmed DNS systems with well-determined total masses  
(uncertainties less than 0.1~$M_\odot$), the PSR J1856--0039 DNS system is the only one with a total mass below $2.5~M_\odot$ (see Fig.~\ref{Mtot-Mns})\footnote{https://www3.mpifr-bonn.mpg.de/staff/pfreire/NS\_masses.html\label{NSmass}}, and its companion is one of the lightest neutron stars known to date,  
consistent with the lower limit of 1.192~$M_\odot$ from simulations of supernova explosions \cite{Muller+2025PhRvL.134g1403M}.

\begin{table*}[htbp]
\centering
\caption{Expected Lense-Thirring precession effects of four DNS systems with the assumption of $I_{\rm p}=10^{45}$~g\,cm$^2$ for all pulsars.}
\setlength{\tabcolsep}{4pt}
\begin{tabular}{lcccccccc}
\hline
PSR name & $\nu$ & $P_{\rm orb}$ & $e$ & $i$ & $\tau_{\rm GW}$ & $\dot{\omega}_{\rm LT}$ & $\dot{x}_{\rm LT}$ & Ref. \\
& (Hz) & (hour) & & (deg) & (Myr) & (deg\,yr$^{-1}$) & ($10^{-14}$ lt-s\,s$^{-1}$) & \\
\hline
J1946+2052 & 58.96   & 1.884 & 0.0638 & $73.7\pm0.3$ & 46 & $-8.84\times10^{-4}$ & $8\sin\delta\sin\Phi_{\rm SO}$ & \cite{Meng+2025AA...704A.153M} \\
\textbf{J1856--0039} & 42.75 & 2.364 & 0.1063  & $46.8\pm1.1$ & 82 & $-4.07\times10^{-4}$ & $10\sin\delta\sin\Phi_{\rm SO}$ & This work \\
J0737--3039A & 44.05 & 2.454 & 0.0878 & $89.35\pm0.05$ & 86 & $-3.77\times10^{-4}$ & $0.18\sin\delta\sin\Phi_{\rm SO}$ & \cite{Kramer+2021PhRvX..11d1050K}\\
J1757--1854 & 46.52  & 4.407 & 0.6058 & $85.3\pm0.2$ & 76 & - & $\lesssim0.3$ & \cite{Cameron+2023MNRAS.523.5064C} \\
\hline
\end{tabular}
\label{LTeffects}
\end{table*}

\section{\label{sec4} Discussions} 

\subsection{Emission geometry and orbital alignment}

Though PSR~J1856--0039 is a recycled pulsar, the polarization angle curve of the averaged polarization profiles in Fig.~\ref{pol} can be well fitted with the rotating vector model \cite[RVM,][]{Radhakrishnan+1969ApL.....3..225R} using \texttt{psrmodel} \cite{Hotan+2004PASA...21..302H}. The model fitting gives the inclination angle of the magnetic axis relative to the spin axis of the neutron star, $\alpha = 114\fdg5 \pm 2\fdg6$, and the inclination angle of the observer's line of sight from the spin axis is $\zeta = 139\fdg1 \pm 3\fdg2$. The best-fitting PA curve is shown by the gray line in Fig.~\ref{pol}(a). 

If there is a large misalignment between the spin axis of the pulsar and the orbital angular momentum of the binary with misalignment angle $\delta$, the geodetic precession will change the angle between the observers' line of sight and the pulsar's spin axis, leading to variations of polarization profiles and position angles in different epochs \cite{Damour+1992PhRvD..45.1840D}. We did not observe such changes over more than 3 years, implying that misalignment is not large. Therefore, we expect that the angle $\zeta$  should be close to the orbital inclination $i$.
However, such a spin-orbit misalignment angle should have a lower limit of $\delta\geq|\zeta-i|$.  From mass measurements and Eq.~\ref{fn}, we found that $i=133\fdg2\pm1\fdg1$. The spin-orbit misalignment angle $\delta$ is thus constrained to $\delta\geq|\zeta-i|=5\fdg9\pm3\fdg4$. 
Note, however, that this value is subject to a large systematic uncertainty from applying the RVM model to millisecond pulsars \cite{Xu+2025A&A...695A.173X}. 

\subsection{Future merger and possible products}

Based on the mass measurement and using equation 5.14 in \cite{Peters+1964PhRv..136.1224P}, we obtained the coalescence timescale $\tau_{\rm GW}$ of PSR~J1856--0039 to be 82~Myr. The product of this DNS merger depends on the mass of its remnant.
As the least massive DNS system known to date (see Fig.~\ref{Mtot-Mns}), if PSR~J1856--0039 is representative of the low-mass end of the DNS merger population, its merger product would be expected to be the lightest one as well. 
To estimate the gravitational mass of the merger remnant, we adopt the relation between gravitational mass $M_{\rm g}$ and baryonic mass $M_{\rm b}$ for a merging-born neutron star proposed by \cite{Gao+2020FrPhy..1524603G}. Because the spin periods of both PSR~J1856--0039 and its companion are much longer than their respective Keplerian periods, the transformation formula for non-rotating neutron stars \cite{Gao+2020FrPhy..1524603G}
\begin{equation}
M_{\rm b}=M_{\rm g}+0.080\,M_{\rm g}^2
\label{mbmg}
\end{equation}
is applicable. Here $M_{\rm b}$ and $M_{\rm g}$ are in units of solar masses. The total merged baryonic mass of the PSR~J1856--0039 binary system is approximately $M_{\rm tot,b}\approx2.74~M_\odot$. During the merger, a fraction of the baryonic mass will be ejected, powering an optical kilonova. The total ejected mass for the GW170817 event is estimated to be $0.07~M_\odot$ \cite[see references therein]{Sarin+2021GReGr..53...59S}. The baryonic mass of the merger remnant would be $M_{\rm rem,b}\approx2.74-0.07=2.67~M_\odot$.

Given this $M_{\rm rem,b}$, one can get a corresponding gravitational mass of $M_{\rm g} \sim 2.26~M_\odot$ for a non-rotating neutron star based on Eq.~\ref{mbmg}. Optical spectroscopy and light-curve modeling have revealed evidence for massive neutron stars in some black widow and redback systems. For instance, \cite{van+2011ApJ...728...95V} reported a pulsar mass of $2.40\pm0.12~M_\odot$ for PSR B1957+20 \cite{Ozel+2016ARA&A..54..401O},  though this is subject to systematic uncertainties from the light-curve modeling. Studies of 18 short $\gamma$-ray bursts also constrain $M_{\rm TOV}$ to $2.31_{-0.21}^{+0.36}~M_\odot$ \cite{Sarin+2020PhRvD.101f3021S}. Considering these observational constraints, the merger remnant of PSR~J1856--0039 will probably become a stable neutron star. Certainly, it is also possible for this merger remnant to collapse into a black hole.

\subsection{Prospects and limitations of the test of GW emission}

To investigate the prospects for testing GW emission, we used the \texttt{fake} plugin in \textsc{Tempo2} \cite{Hobbs+2006MNRAS.369..655H} to simulate future timing of this pulsar, assuming a timing precision of 5.8\,\textmu s (equal to the current $\sigma_{\rm res}$). Assuming a 2-hour tracking observation every 60 days, a total time span of 7000 days, and a 5-minute integration time per TOA, we obtained the simulated TOAs. Using these TOAs and the \texttt{DD} model \cite{Damour+1985AIHPA..43..107D,Damour+1986AIHS...44..263D}, the uncertainties of the PK parameters are constrained to $\sigma_{\dot\omega}=2\times10^{-5}$ deg\,yr$^{-1}$, $\sigma_{\dot{P}_{\rm orb}}=2\times10^{-16}$ s\,s$^{-1}$ and $\sigma_{\gamma}=2\times10^{-4}$ ms.

The proper motion of the pulsar changes the vector of the line of sight, leading to a difference between the observed and intrinsic $\dot\omega$ \cite{Kopeikin+1996ApJ...467L..93K}:
\begin{equation}
    \dot\omega^{\rm PM}=\csc i(\mu_{\alpha}\cos\Omega_{\rm asc}+\mu_{\delta}\sin\Omega_{\rm asc}),
    \label{eq:omdotpm}
\end{equation}
where $\Omega_{\rm asc}$ is the position angle of the ascending node. According to Eq.~\ref{eq:omdotpm}, $|\dot\omega^{\rm PM}|\leq\csc i\sqrt{\mu_{\alpha}^2+\mu_{\delta}^2}\approx2\times10^{-6}$ deg\,yr$^{-1}$, which is much smaller than $\sigma_{\dot\omega}$. Therefore, the difference between the observed and intrinsic values of $\dot\omega$ can be neglected. By combining $\dot\omega$ and $\gamma$, we can obtain values of $M_{\rm p}$ and $M_{\rm c}$ and therefore make a precise prediction $\dot{P}_{\rm orb,pred}^{\rm GR}$. The total kinematic correction to $\dot{P}_{\rm orb}$, depending on the distance accuracy of the pulsar, is approximately $1\times10^{-15}$ s\,s$^{-1}$. 
Owing to the low flux density of PSR~J1856--0039, it is hard to measure the pulsar parallax through Very Long Baseline Interferometry (VLBI) observations. Long-term timing may be able to measure its parallax. 
Another approach is to observe such nearby DNS systems by 
future space-based GW detectors in the sub-mHz band \cite{Chen+2026PhRvD.113h4013C}, including LISAmax \cite{Martens+2023CQGra..40s5022M}, Folkner \cite{Mueller+2019BAAS}, and eASTROD \cite{Ni+2013IJMPD..2241004N}, greatly improving the precision of gravity tests based on $\dot{P}^{\rm GW}_{\rm orb,obs}$.

\subsection{Probably measurable Lense-Thirring precession}

General relativity predicts that rotating massive bodies can drag spacetime around them, affecting the motion of nearby objects. This relativistic frame-dragging effect leads to a secular precession of orbits, known as ``Lense-Thirring precession". Measuring Lense-Thirring precession in DNS systems allows one to infer the moment of inertia of the neutron star, thereby providing a novel constraint on the dense-matter equation of state. In pulsar timing, the Lense-Thirring effect manifests as two observable contributions: an additional periastron advance $\dot{\omega}_{\rm LT}$ and a secular change in the projected semi-major axis $\dot{x}_{\rm LT}$. Assuming that the spin-orbit misalignment angle is small and that the companion star rotates slowly, general relativity gives \cite{Hu+2024Univ...10..160H}:
\begin{equation}
\begin{cases}
\dot{\omega}_{\rm LT}=-\left(\dfrac{2\pi}{cP_{\rm orb}}\right)^{2}\dfrac{4M_{\rm p}+3M_{\rm c}}{M_{\rm p}(M_{\rm p}+M_{\rm c})}\dfrac{2\pi\nu I_{\rm p}}{(1-e^2)^{3/2}},\\[2.5ex]
\dot{x}_{\rm LT}=-\dfrac{1}{2}x\cot i\sin\delta\sin\Phi_{\rm SO}\times\dot{\omega}_{\rm LT},
\end{cases}
\end{equation}
where $\nu$ is the pulsar spin frequency, $I_{\rm p}$ is its moment of inertia, and $\Phi_{\rm SO}$ is the geodetic precession phase as defined by \cite{Cameron+2023MNRAS.523.5064C}. $\dot\omega_{\rm LT}$ and $\dot{x}_{\rm LT}$ can be used to constrain $I_{\rm p}$, and a low orbital inclination and a large spin-orbit misalignment can make the measurement of $\dot{x}_{\rm LT}$ easier.

Attempts to measure the NS-spin-induced Lense-Thirring precession, via either $\dot{\omega}_{\rm LT}$ or $\dot{x}_{\rm LT}$, have been made in the double pulsar, PSR J0737--3039A \cite{Kramer+2021PhRvX..11d1050K} and the most accelerated binary pulsar, PSR J1757--1854 \cite{Cameron+2023MNRAS.523.5064C}. 
The most compact DNS system, PSR J1946+2052, also shows great potential for measuring the Lense-Thirring precession \cite{Meng+2025AA...704A.153M}. The expected Lense-Thirring precession amplitudes for these systems are summarized in Table~\ref{LTeffects}. Among them, the spin-orbit misalignment of PSR J1757--1854 is significant \cite{Cameron+2023MNRAS.523.5064C}, while those of PSRs J0737--3039A and J1946+2052 are small. PSR J0737--3039A ($e=0.0878$) has $\delta < 3\fdg2$ \cite{Ferdman+2013ApJ...767...85F}; and PSR J1946+2052 ($e=0.0638$) has $\delta = 0\fdg21^{+0\fdg28}_{-0\fdg10}$ \cite{Meng+2024ApJ...966...46M}. The small misalignment angles make detecting $\dot{x}_{\rm LT}$ difficult. The large misalignment angle of PSR~J1757--1854 ($e=0.6058$) makes it a good candidate to measure $\dot x_{\rm LT}$ \cite{Cameron+2023MNRAS.523.5064C}. For PSR J1856--0039, its $\dot{x}_{\rm LT}$ is possibly comparable to or larger than that of PSR~J1757--1854 if the $\delta$ is really a few degrees. The $\dot{\omega}_{\rm LT}$ of PSR~J1856--0039 is also comparable to those of PSRs J0737--3039A and J1946+2052. With its short orbital period and low orbital inclination, PSR~J1856--0039 emerges as a good candidate for future measurements of Lense-Thirring precession.

Based on the simulated TOAs with a 7000-day time span and using the DDGR timing model \cite{Taylor+1987grg..conf..209T, Taylor+1989ApJ...345..434T}, we found that the uncertainty of $\dot{x}$ is about $1\times10^{-15}$ lt-s\,s$^{-1}$. Besides orbital precession due to LT effects, another important contribution to $\dot{x}$ from the proper motion is \cite{Kopeikin+1996ApJ...467L..93K}
\begin{equation}
    \frac{\dot{x}_{\rm PM}}{x}=\cot i(-\mu_{\alpha}\sin\Omega_{\rm asc}+\mu_{\delta}\cos\Omega_{\rm asc}).
\end{equation}
$|\dot x_{\rm PM}|$ is also about $1\times10^{-15}$ lt-s\,s$^{-1}$. Because $\delta$ is small, changes of $x$ due to varying aberration can be neglected \cite{Damour+1992PhRvD..45.1840D}. Compared with $\dot{x}_{\rm LT}=(\delta/1^\circ)\sin\Phi_{\rm SO}\times1.7\times10^{-15}$ lt-s\,s$^{-1}$, $\dot{x}_{\rm LT}$ could possibly be detected if $\delta$ is only a few degrees.

To measure $\dot\omega_{\rm LT}$, one has to use $\dot{P}^{\rm GW}_{\rm orb,obs}$ and $\gamma$ to extract $\dot\omega_{\rm LT}$ from the total $\dot\omega$. Because $M_{\rm p}+M_{\rm c}\gg M_{\rm p}-M_{\rm c}$, in general relativity, $\dot P^{\rm GW}_{\rm orb,pred}$ can be approximated as
\begin{equation}
    \dot{P}^{\rm GW}_{\rm orb,pred}\approx-\dfrac{48\pi}{5}\left(\dfrac{2\pi G}{c^3 P_{\rm orb}}\right)^{5/3}(M_{\rm p}+M_{\rm c})^{5/3}f(e).
\end{equation}
Therefore, assuming that $\dot{P}^{\rm GW}_{\rm orb,obs}=\dot{P}^{\rm GW}_{\rm orb,pred}$, one then finds
\begin{equation}
    -\frac{\sigma_{\dot{P}^{\rm GW}_{\rm orb,obs}}}{\dot{P}^{\rm GW}_{\rm orb,obs}}\approx\frac{5}{3}\frac{\sigma_{\rm M_{\rm tot}}}{M_{\rm tot}}=\frac{5}{2}\frac{\sigma_{\dot\omega_{\rm LT}}}{\dot\omega}.
\end{equation}
To obtain a reliable measurement of $\dot\omega_{\rm LT}$, the uncertainty $\sigma_{\dot{P}^{\rm GW}_{\rm orb,obs}}$ must satisfy $\sigma_{\dot{P}^{\rm GW}_{\rm orb,obs}}\ll7\times10^{-17}$ s\,s$^{-1}$. Achieving such precision requires an observing time span of about half a century and an independent distance measurement of the binary. \\[5mm]

\begin{acknowledgments}
We sincerely thank the three reviewers for their very careful reading of the manuscript and very thoughtful suggestions. The authors were supported by the National Key R\&D Program of China No. 2025YFA1614002, the National Natural Science Foundation of China (NSFC), grant numbers 12588202 and 12041303, the Chinese Academy of Sciences via project JZHKYPT-2021-06, 
%
%
and the National SKA Program of China grant 2020SKA0120100 and 2022SKA0120103.
This work made use of the data from FAST (https://cstr.cn/31116.02.FAST): the first 7 sessions observed by the projects: ZD2020\_2, ZD2021\_2 and ZD2022\_2 (PI: J.~L. Han), the later 9 sessions by PT2023\_0143 and PT2024\_0020 (PI: Z.~L. Yang) and one session in the middle by PT2023\_0102 (PI: W.~Q. Su). FAST is a Chinese national mega-science facility, operated by the National Astronomical Observatories, Chinese Academy of Sciences. 
\end{acknowledgments}


\bibliography{LR20337.bib}


\end{document}